\documentclass[twocolumn,journal]{IEEEtran}
\usepackage[T1]{fontenc}
\usepackage[latin9]{inputenc}
\usepackage{graphicx}
\usepackage[unicode=true,
 bookmarks=true,bookmarksnumbered=true,bookmarksopen=true,bookmarksopenlevel=1,
 breaklinks=false,pdfborder={0 0 0},pdfborderstyle={},backref=false,colorlinks=false]
 {hyperref}
\hypersetup{pdftitle={Your Title},
 pdfauthor={Your Name},
 pdfpagelayout=OneColumn, pdfnewwindow=true, pdfstartview=XYZ, plainpages=false}

\makeatletter
 \let\oldforeign@language\foreign@language
 \DeclareRobustCommand{\foreign@language}[1]{%
   \lowercase{\oldforeign@language{#1}}}

\usepackage[caption=false,font=footnotesize]{subfig}
\usepackage{flushend}

\@ifundefined{showcaptionsetup}{}{%
 \PassOptionsToPackage{caption=false}{subfig}}
\usepackage{subfig}
\makeatother

\begin{document}

\title{Reversible Hardware for Acoustic Communications}

\author{Harun Siljak,~\IEEEmembership{Member,~IEEE,} Julien de Rosny,~\IEEEmembership{Member, IEEE}
and~Mathias Fink\thanks{This publication has emanated from research supported in part by a
research grant from Science Foundation Ireland (SFI) and is co-funded
under the European Regional Development Fund under Grant Number 13/RC/2077.
The project has received funding from the European Unions Horizon
2020 research and innovation programme under the Marie Skodowska-Curie
grant agreement No 713567 and was partially supported by the COST
Action IC1405.}\thanks{Harun Siljak is with CONNECT Centre, Trinity College Dublin, Ireland,
e-mail: siljakh@tcd.ie.}\thanks{Julien de Rosny and Mathias Fink are with Institut Langevin, ESPCI
ParisTech, CNRS UMR 7587, 75231 Paris Cedex 05, France, e-mails: \{julien.derosny,mathias.fink\}@espci.fr.}}

\markboth{}{Siljak \MakeLowercase{\emph{et al.}}: Reversible Hardware for Acoustic
Communications}
\maketitle
\begin{abstract}
Reversible computation has been recognised as a potential solution
to the technological bottleneck in the future of computing machinery.
Rolf Landauer determined the lower limit for power dissipation in
computation and noted that dissipation happens when information is
lost, i.e., when a bit is erased. This meant that reversible computation,
conserving information conserves energy as well, and as such can operate
on arbitrarily small power. There were only a few applications and
use cases of reversible computing hardware. Here we present a novel
reversible computation architecture for time reversal of waves, with
an application to sound wave communications. This energy efficient
design is also a natural one, and it allows the use of the same hardware
for transmission and reception at the time reversal mirror.
\end{abstract}


\section{Introduction}

\IEEEPARstart{T}{he} majority of computation we perform is irreversible:
addition of two numbers, or logical AND of two bits both destroy the
information about the inputs. In a computation paradigm which prioritises
saving memory resources, losing information is a consequence of using
a register for something else as soon as its current content is used
for the last time. The arithmetic units are often designed so that
the result replaces one of the input operands, and this is considered
an important save in resources. Somewhat paradoxically, the call for
reversibility and preservation of information through the computation
process also comes from the resource optimization perspective.

Thermodynamics of computation explains the mechanisms of energy use
and dissipation in computing systems. Landauer \cite{landauer_irreversibility_1961}
established an important lower limit for computation energy dissipation:
the erasure of one bit takes a fraction of a joule, energy proportional
to the working temperature of the system (with the proportionality
coefficient equal to the product of the Boltzmann constant and natural
logarithm of two). This lower limit follows from the equivalence of
thermodynamical and informational entropy and, at the time, it was
significantly lower than the limits imposed by technological (semiconductor)
constraints. With the advancement of semiconductor industry, the limits
of devices became closer to Landauer\textquoteright s limit. Landauer\textquoteright s
limit bounds bit erasure: operations that do not erase information
(bits) do not have a lower bound in thermodynamical-informational
sense. This fact qualifies information-conserving computation as a
potential solution for the future of general computing in the post-Moore
law era.

The concept of reversible computation, reversible logic gates and
circuit design have been a topic of research since Bennett's pioneering
work \cite{bennett_logical_1973} on applying Landauer's ideas to
hardware. However, there have been almost no applications of reversible
circuits to real world problems, no interfaces with the nature and
other technology. In this paper, we present a case for employing reversible
computation in wave time reversal, using acoustic underwater communication
as a working example.

The case of acoustic communications based on wave time reversal is
a good ground for reversible computation. Everything is reversible:
the communication scheme reversing the carrier wave toward the original
source and the environment obeying reversible Euler equation. Here
we show how the computation performing all of it can be reversible
as well. Of course, wave time reversal is not limited to acoustic
communication, as it is the basis of a beamforming approach for RF
communications. Hence, our contribution is relevant to multiple wireless
communications paradigms. Our motivation for presenting the reversible
hardware solution for wave time reversal as a contribution to communications
stems from here: a solution for time-reversal massive MIMO would be
an adaptation of the one presented here, and same holds for optical
communications based on time reversal. In this manner, digital signal
processing in communications would be ready for the post-Moore age
of reversible and/or quantum computing. Time-reversal based communications
implemented with this circuitry then have both the potential of immense
energy efficiency and a chance to become the natural solution for
the physical layer of quantum networks of the future.

We begin by revisiting the mechanism of wave time reversal, followed
by a presentation of reversible computation and motivation of its
use. We discuss the possible design options employing reversible hardware
for time reversal and show design results. We conclude with a discussion
of the proposed solution, future research and the wider effect of
reversible hardware introduction in wireless communications.

\section{Reversibility of waves and computation}

It is not a coincidence that we chose wave time reversal for the demonstration
of an efficient reversible hardware application. As this section will
show, the wave time reversal and reversible computation both rely
on keeping the information about backtracking known, to run backwards.
They share the same philosophy and a common foe.

\begin{figure}[htbp]
\begin{centering}
\textsf{\includegraphics[width=0.8\columnwidth]{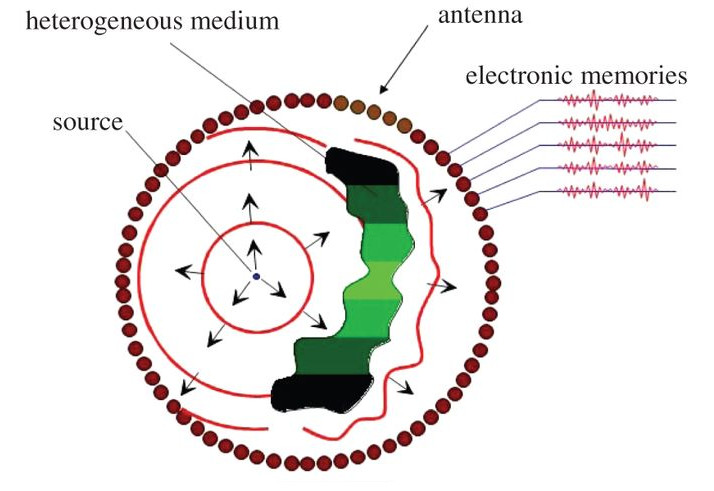}}
\par\end{centering}
\caption{The wavefront distorted by heterogeneities comes from a point source
and is recorded on the cavity elements. In the next step the recorded
signals are time-reversed and re-emitted by the elements. The time-reversed
field back-propagates and refocuses exactly on the initial source.\cite{fink2016loschmidt}}
\label{trm}
\end{figure}

\subsection{Wave Time Reversal}

Both, reversing waves and reversing computation are plagued with the
scale destroying reversibility. While at the microscale the elementary
components of the system are obeying reversible laws (be it simple
computational operations, be it equations of motion), their ensembles
lose the reversibility at the macroscale. Loschmidt\textquoteright s
thought experiment with a deamon capable of reversing all velocities
of particles in a gas and hence reversing the behaviour of the ensemble
asks for too much information and ability on the deamon\textquoteright s
side, but it is a worthy goal to pursue: how can we reverse a propagating
wave so it ends up converging at its original source?

The solution based on time reversal mirrors (TRMs) \cite{fink_time_1992}
performs this regardless of the complexity of the medium as if time
were going backwards, and has been implemented with acoustic, electromagnetic
and water waves. It requires the use of emitter\textendash receptor
antennas positioned on an arbitrary enclosing surface. The wave is
recorded, digitized, stored, time-reversed and rebroadcasted by the
same antenna array. If the array intercepts the entire forward wave
with a good spatial sampling, it generates a perfect backward-propagating
copy. 

The principle of wave time reversal builds upon the exact nature of
the wave equation, and its solution being a continuous function of
three spatial and one temporal dimension, i.e. described over a hypervolume
with four variables, bounded by a hypersurface with three variables.
This boundary can either be observed as composed of three spatial
dimensions or of two spatial and one temporal dimension . Depending
on the choice of the boundary description, we have two different approaches
to time reversal, dubbed \textquotedblleft à la Huygens\textquotedblright{}
(named after Huygens integral theorem) and \textquotedblleft à la
Loschmidt\textquotedblright{} (named after Loschmidt\textquoteright s
deamon). \cite{fink2016loschmidt}

\subsubsection{The time-reversal mirror approach \textquoteleft à la Huygens\textquoteright{}}

In this approach, represented in the Fig. \ref{trm}, a transient
wavefield originating from the initial source is radiated throughout
a heterogeneous medium closed in a cavity bounded by a two-dimensional
surface. This surface is populated with sensing and recording devices
keeping the information about the wavefield and its normal derivative.
This process continues until the incoming field vanishes along the
boundary. This recording suffices for the recovery of the wavefield,
as we will soon see. 

Out of the two solutions of the wave equation (wave operator obeys
time-reversal symmetry), the causal one is radiated from the source,
and we aim to radiate the anti-causal one from the boundary. The collected
samples are hence time-reversed and rebroadcasted by the same antenna
array that has collected them.

This new wave satisfies a homogeneous wave equation with the time-reversed
boundary conditions without the original source. Hence it is not enough
to time-reverse the wavefield on the boundary, as the original source
needs to be reversed into a sink. While returning to the original
source, the re-emitted wave does appear to converge, but as it cannot
stop on its own, after the collapse it continues to propagate in divergent
manner. To compensate this diverging field, we either use an active
source at the focusing point canceling the field, or a passive sink
as a perfect absorber. \cite{de2002overcoming}

\begin{figure}[tbh]
\begin{centering}
\includegraphics[width=0.8\columnwidth]{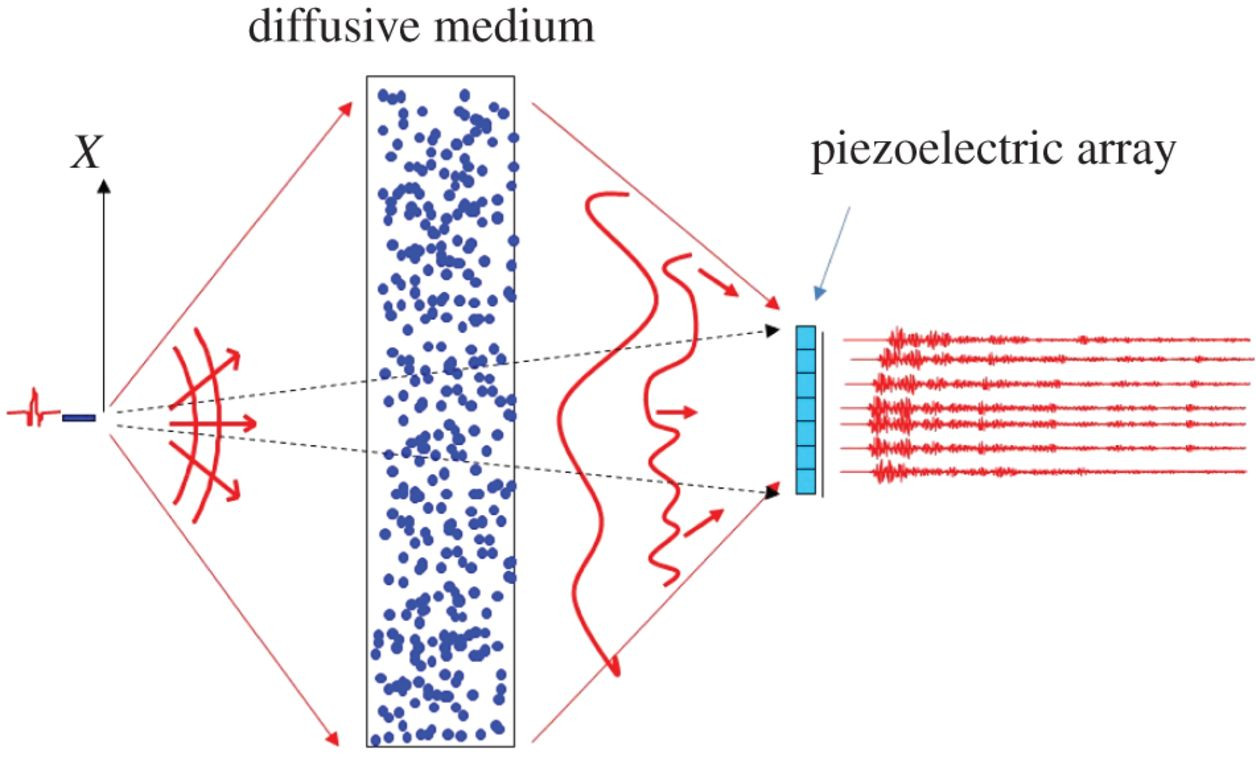}
\par\end{centering}
\caption{Time-reversal experiment through a diffusive medium \cite{fink2016loschmidt}}

\label{diffus}
\end{figure}

So far, we assumed the idealised case where the entire surface of
the boundary is covered with transceivers, which requires a large
number of hardware components. This requirement can be dropped, and
the execution of the reversal can be simplified. One way is locating
the TRM in the far field of the source and of the medium heterogeneities.
This halves the quantity of data to be stored, as the normal derivative
remains proportional to the field and does not have to be recorded
at all. In addition to this, it was experimentally shown that TRM
consisting of a small number of elements (time-reversal channels)
functions on a limited angular area as well, when it uses complex
environments to appear wider than it is. The resulting refocusing
quality does not depend on the TRM aperture. In that regard, observe
the following experiment setup.\cite{derode1995robust} A point-like
transducer is separated from a TRM by a large distance (much larger
than the wavelength) and by a multiple scattering medium (forest of
steel rods) and shown in Fig. \ref{diffus}. After emitting a short
pulse from the source, the sensors at the TRM collect the impulse
response. The spread of these impulse responses is two orders of magnitude
higer than the initial pulse duration as the multi-scattering medium
is highly diffusive. As explained before, in the next step the responses
are flipped in memory and re-transmitted from the TRM. The impulse
duration, reconstructed at the original source is the same as the
original; the spatial spread is a level of magnitude smaller (i.e.
more coherent) when the propagation happens in the complex medium,
than in the case of free space propagation.

\subsubsection{The instantaneous time mirror approach \textquoteleft à la Loschmidt\textquoteright{}}

Going back to Loschmidt\textquoteright s demon able to turn the direction
of particles instanteneously, if one decides to imitate it, unlike
the Huygens case, the measurements of the incoming wavefield have
to be performed at one specific time in the whole volume. At that
point a new set of initial conditions is imposed, in which the sign
of the time derivative is reversed, and the resulting wave is the
time-reversed original.

Examine a case of a bath of fluid, placed on a shaker to control its
vertical motion. After emitting a pulse from a point on the fluid
surface, at the chosen time instant a vertical downwards acceleration
is applied to the bath, an impulsive change of wave celerity which
can be described by a delta function in time. While the propagation
of the initial outwards propagating wave is not affected, a new contribution
emerges: a backwards converging circular wave packet. Just like in
the Huygens case, this wave packet focuses at the original source
and proceeds diverging afterwards. While the result is the same, we
note that in this case no transceivers or memory elements were used:
the information is stored in the medium itself.

\subsection*{The Bridge to Computing}

Among time reversal studies that have followed the development of
these concepts one creates a link between computation and time reversal
in waves \cite{perrard_wave-based_2016}. A dissipative chaotic system
consisting of a drop bouncing on a vibrated liquid bath, exchanging
information with the waves it forms, can be reversed. The elementary
motions performed by the system are equivalent to writing, storing,
reading and erasing operations of a Turing machine. The bouncing drop
reads information as it backtracks, at the same time it is erasing
the read information. In the next section, we investigate computational
systems which use an equivalent principle to perform useful calculations
and save power.

\subsection{Reversible Hardware \label{subsec:Reversible-Hardware}}

Landauer famously concluded that ``the information is physical''
and brought together Shannon's and Boltzmann's views on entropy. \cite{landauer_irreversibility_1961}
Digital computation that does not lose information (erase bits of
information), does not have to dissipate power. When we delete a bit,
the information that was stored there physically moves to the environment
in form of heat, a direct display of Boltzmann\textquoteright s thermodinamical
entropy. None of it would have to be dissipated from the entropical
perspective if the erasure was not performed.

Observe a digital circuit consisting of logical gates: e.g. a single
AND gate with its two inputs and one output. Its output is one when
both inputs are one simultaneously; otherwise it is zero. Hence the
knowledge about the output is not enough to tell us the inputs, as
three different input combinations collapse into one output state.
If we want to make an information-preserving gate, it has to have
a one-to-one correspondence between output states and input states.
This asks for the same number of outputs and inputs in such \emph{reversible
gates}.

\begin{figure}[tp]
\begin{centering}
\includegraphics[width=0.8\columnwidth]{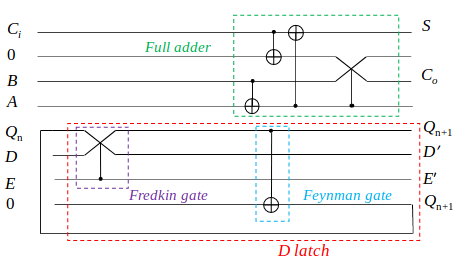}
\par\end{centering}
\caption{Reversible circuits: D-latch, full adder, and logical gates as building
blocks}
\label{gates}
\end{figure}

Two reversible bit operations are bit inversion and swap of two variables.
To make more use of them, we devise gates controlling these operations
according to the state of other variables. Fig. \ref{gates} shows
some basic reversible gates as part of larger reversible circuits:
all circuits in the figure are built using Feynman and Fredkin gates.
Feynman gate is a controlled NOT: the variable with the $\oplus$
is inverted if and only if the control input, the variable with $\bullet$
is equal to 1. In a more general setting of the Toffoli gate, multiple
variables can control a single NOT; in that case the function controlling
the gate is an AND of the control inputs. Similarly, the Fredkin gate
swaps the variables joining in the $\times$ if and only if the variable(s)
with $\bullet$ are equal to 1.

These gates enable design of reversible circuits which perform the
usual digital electronics tasks. A full adder \cite{skoneczny_reversible_2008}
and D-latch \cite{nayeem_efficient_2009} are shown in Fig. \ref{gates}.
Additional inputs/outputs are auxilliary variables\textendash the
\emph{ancilla bits}. In the D-latch example, the 0 bit and the Feynman
gate attached to it are necessary to copy the latch output for feedback,
as reversible circuits do not allow fan-out (it violates the one-to-one
correspondence requirement).

Most classical computation is irreversible, re-using the memory by
often removing intermediate results. In the past, most of the dissipation
in logic circuits came from imperfections of the practical implementation.
With the progress of semiconductor technology, the dissipation levels
are approaching those of Landauer\textquoteright s limit, and reversibility
is gaining importance. As Moore\textquoteright s law comes to its
potential end, and alternative solutions are sought, one of the candidates
being reversible computing. Irreversible calculations could be easily
embedded in a reversible computation by merely keeping track of the
steps made. To mitigate the need for unbounded memory, Bennett \cite{bennett_logical_1973}
introduced a trick: if a computation is made in a reversible circuit
in one direction and then in reverse (computed, and then uncomputed),
the memory occupied in the direct pass is freed in the return pass,
and it is available for a new computation without bit erasures, and
entropy is not increased (the memory after the return pass is back
to the state before the direct pass).

The lack of application for reversible computation in the classical
realm is reversed in the quantum computing domain. Most quantum computing
schemes require reversibility to operate, so the reversible logic
gates are constituent parts of quantum circuits (and are often referred
to as \emph{quantum gates}). Reversible computation is not limited
to electronics (where \emph{adiabatic circuits} are known for several
decades \cite{frank2017asynchronous}) and quantum computers: (micro)electromechanical
systems and quantum dots are also capable of reversible computation.

\section{The Design}

\begin{figure}[tbh]
\begin{centering}
\includegraphics[width=0.8\columnwidth]{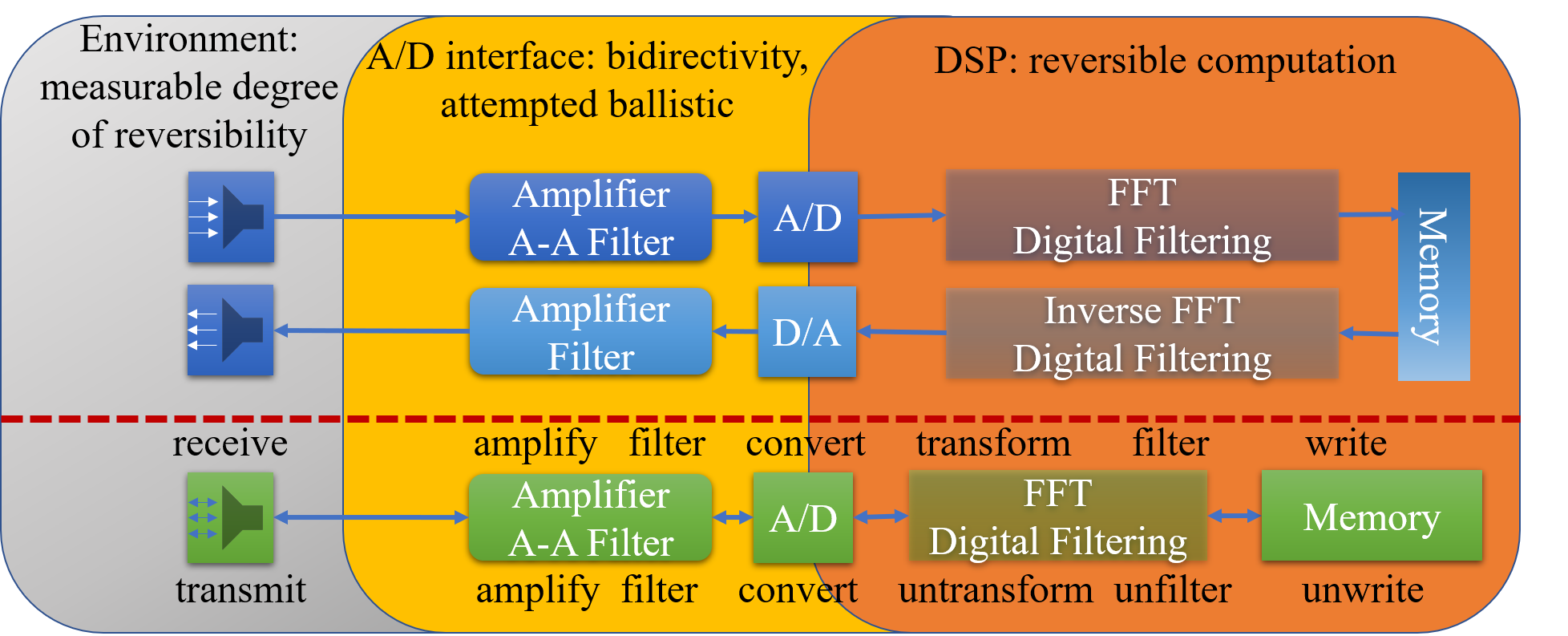}
\par\end{centering}
\caption{The classical (top) and the reversible solution (bottom) for the classical
time reversal chain.}
\label{conceptual}
\end{figure}

We have seen so far a computational paradigm relying on reversibility
of calculation, and a communication scheme relying on the reversibility
of wave propagation. In this part of the article, we proceed with
designing a TRM based on reversible hardware. The reversible gates
will form the digital logic part in the TRM, but the design is more
complex than just digital logic. With Fig. \ref{conceptual} we illustrate
the layers of the design task:
\begin{enumerate}
\item The environment is reversible to an extent. In the use case of acoustic
underwater communication, the physics of wave propagation in water
is reversible, but the issues arise as we lose information in the
process: parts of the wave might end up reflected to unreachable parts
of the environment if the observed space is not ergodic, guaranteeing
all parts of the environment to be visited by the wave components.
The hardware in contact with the environment are the microphones and
speakers, i.e. sensors and actuators.
\item The analog computation part of the TRM loses information. It comprises
of anti-alias filters before analog-to-digital conversion (ADC), filters
after digital-to-analog conversion (DAC), amplifiers accompanying
the filters and the converters themselves, at the transition to the
digital domain. We analyse these components in Section \ref{subsec:Analog-processing}.
\item Finally, the digital computation part of the TRM is reversible and
no increase in entropy is necessary. This part entails writing in
memory and unwriting, in the fashion of Bennett\textquoteright s trick,
enabling reuse of memory for the next incoming wave, while not increasing
the entropy. It may include a transform into frequency domain and
digital filtering, and we discuss these options as well in Section
\ref{subsec:Digital-processing}.
\end{enumerate}

\subsection{Analog processing\label{subsec:Analog-processing}}

Analog processing is the lossy, inherently irreversible bridge between
two domains which exhibit reversibility: the physics of wave propagation
and the digital signal processing chain we introduce. Our main goal
at this point is to make the mechanisms in this part of the processing
chain bidirectional, so that they can be used both for the inputs
and the outputs. However, we are interested in saving as much information
as we can, so we investigate the information loss in this part of
the chain as well.

\subsubsection{Bidirectional amplification and AD/DA conversion}

From the information-preserving perspective, ideal amplification is
not an interesting process. However, the real amplifier is an imperfect
device with a limited bandwidth and it loses signal information and
introduces changes in the signal shape. It requires additional energy
for the signal, hence we have to allocate a non-zero energy budget
for this part of the computation. At the same time, the analog to
digital and digital to analog converters both modify the signal they
convert due to finite resolution and sampling rates, losing information
about the original signal. However, the idea of the single device
performing as both an ADC and a DAC depending on the direction exists
both in academia and industry, with a large number of patents describing
these bi-directional devices. \cite{suarez1975reversible} In our
proposed solution, we assume that the bi-directional converters are
bundled with bi-directional amplifiers \cite{azadmehr2009bi}. We
note the complexity of structure and switching in these devices.

\subsubsection{The information-increasing filter}

A part of analog to digital conversion is the anti-aliasing filter.
Filtering, in the most common interpretation, removes a part of the
signal and hence loses information about the original signal. However,
the anti-aliasing filter is employed to prevent significant information
loss due to spectrum overlaps in the analog-to-digital conversion,
and hence in this situation represents an information gain and may
be re-interpreted in the context of \emph{useful, relevant information}
\cite{geiger_information_2018}. To avoid confusion, we may consider
the anti-aliasing filter as a part of the analog-to-digital converter
and as such, implement it in the bi-directional fashion.

\subsubsection{One-bit reversal}

The conversion is additionally simplified in the one-bit solution
\cite{derode_ultrasonic_1999} where the receivers at the mirror register
only the sign of the waveform and the transmitters emit the reversed
version based on this information. It is a special case of analog-to-digital
and digital-to-analog conversion with single bit converters. The reduction
in discretisation levels also means simplification of the processing
chain and making its reversal (bi-directivity) even simpler. The question
of the information loss is not straightforward: while the information
about the incoming wave is lost in the conversion process (and the
loss is maximal due to minimal resolution), spatial and temporal resolution
are not significantly degraded.

This scheme can also be called \textquotedblleft one-trit\textquotedblright{}
reversal: there are three possible states in the practical implementation:
positive pressure, negative pressure, and \textquoteright off\textquoteright .
Reversibility and multi-valued logic were going hand in hand from
the beginning: binary reversible logic is just a special case of multi-valued
reversible logic. Hence, this scheme is readily implementable in reversible
logic as well.

\subsection{Digital processing\label{subsec:Digital-processing}}

With the functionalities reversible gates presented in Section \ref{subsec:Reversible-Hardware}
can offer when combined into logical circuits, building a digital
signal processing chain for wave time reversal becomes the matter
of combining circuits into more complex structures, akin to traditional
circuit design. The idea we show here can easily be translated into
any scheme of modulation-demodulation, coding-decoding, which are
often seen in communications hardware and software. While fully functional
and directly applicable, our time reversal signal processing chain
is a proof of concept for reversible communications signal processing
of arbitrary complexity.

\subsubsection{Time domain reversal}

The first, straightforward way of performing time reversal of a digitally
sampled wave is storing it in memory and reading the samples in the
reverse order (last in, first out, LIFO), analoguous to storing the
samples on the stack. The design of registers in reversible logic
is a well-explored topic \cite{nayeem_efficient_2009} and both serial
and parallel reading/writing can be implemented. In the sense of already
presented circuits, we have seen a design for the D-latch (Fig. \ref{gates}):
a combination of latches makes a flip-flop, and a series of flip flops
makes a register (and a reversible address counter). In the case of
wave time reversal, this is important to know, as the two possible
variants of wave time reversal can be interpreted as two variants
of memory writing:
\begin{enumerate}
\item Wave reversal à la Loschmidt is a large register being loaded in parallel
with wave data;
\item Wave reversal à la Huygens is a large register being loaded serially
with wave data.
\end{enumerate}
In the case of a localized time reversal mirror (all samples at the
same place) $m$ bits from the ADC are memorised at the converter's
sample rate inside a $k\times m$ bit register matrix (where $k$
is the number of samples to be stored for time reversal). In the receiving
process, the bits are stored, in the transmission process they are
\emph{unstored}, returning the memory into the blank state it started
from (uncomputation). We utilise Bennett's trick and lose information
without the entropic penalty: the information is kept as long as it
is relevant.

\subsubsection{Frequency domain reversal}

When additional signal processing, e.g. filtering or modulation is
performed, it is convenient to reverse waves in frequency domain:
there, time domain reversal is achieved by phase conjugation, i.e.
changing the sign of the signal's phase. The transition from time
to frequency domain (and vice versa) in digital domain is performed
by the Fast Fourier Transform (FFT) and its inverse counterpart. These
procedures are inherently reversible and information-preserving, and
their implementation in reversible digital circuits asks for a network
of reversible adders (and reversible multipliers, again comprised
of adders) \cite{skoneczny_reversible_2008}, and we have already
seen an implementation of a reversible adder in Fig. \ref{gates}.
The necessary phase conjugation is an arithmetic operation of sign
reversal, again perfectly reversible and with a known implementation.
The additional signal processing can be performed reversibly as well:
one example is the filtering process done through filter banks and
wavelet transforms. It remains reversible as all components of signals
are preserved, if nothing then as the remainder \cite{suzuki2018redefined}.
Both the reversible wavelet computation and reversible Fast Fourier
Transform use the lifting scheme.

\begin{figure}[htbp]
\begin{centering}
\subfloat[]{\begin{centering}
\includegraphics[width=0.8\columnwidth]{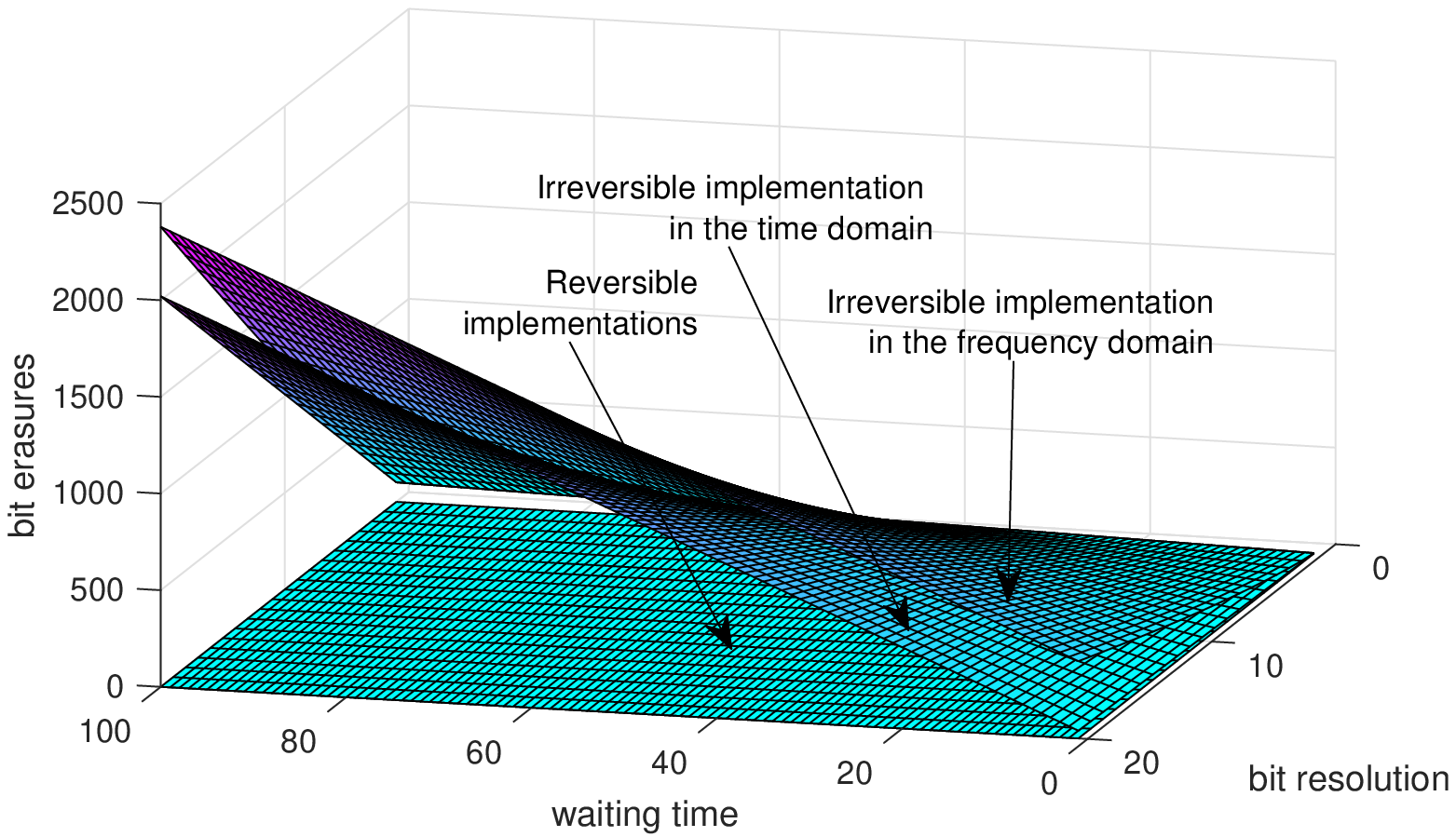}
\par\end{centering}

}
\par\end{centering}
\begin{centering}
\subfloat[]{\begin{centering}
\includegraphics[width=0.8\columnwidth]{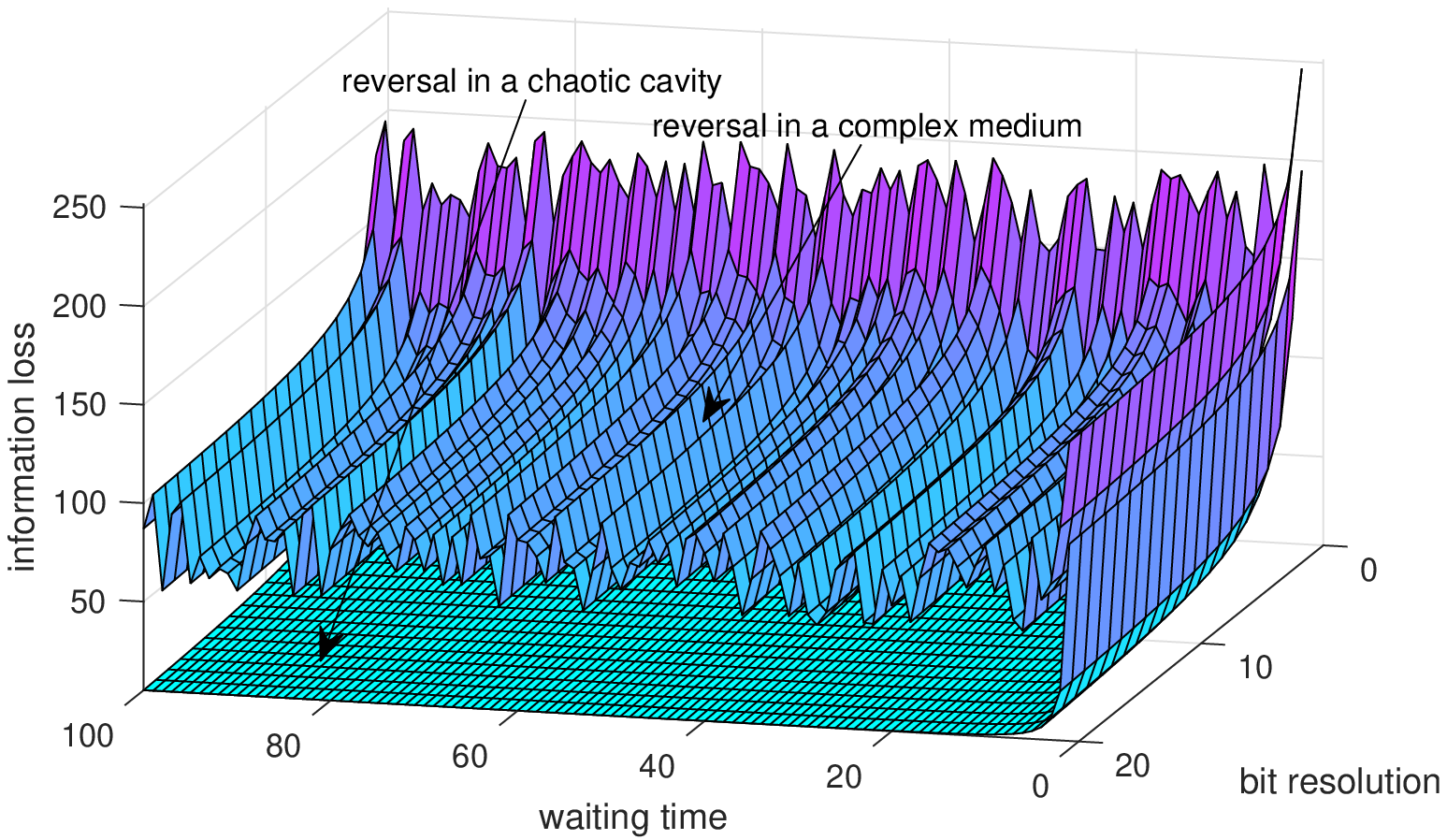}
\par\end{centering}
}
\par\end{centering}
\caption{Information loss in (a) digital and (b) analog part of the system.
Units are omitted as the particular aspects of implementation are
not relevant for the illustration of effects. Plot (a) is obtained
by counting operations, plot (b) by simulation of back-scattering,
both originating from theoretical calculations.}
\label{graphs}
\end{figure}

\section{Discussion and Conclusions}

We have presented the components to be used in the implementation:
where possible (in the digital domain) we use reversible circuits,
otherwise we use bidirectional components. What is the gain of the
new implementation? As already suggested, the loss of information
is directly related to the dissipation of energy in the system, so
let us observe how does our solution fare in this regard.

Fig. \ref{graphs}(a) is a comparison of the bit erasures in different
implementations of the digital circuitry: frequency domain (FFT) and
time domain reversal performed by irreversible circuits, compared
to reversible implementations. The number of erasures changes depending
on two parameters: bit resolution of the ADC and the waiting time\textendash the
length of the interval in which samples are collected before reversal
starts, equivalent to the number of digitised samples. The increase
in both means additional memory locations and additional dissipation
for irreversible circuits. The irreversible FFT implementation has
an additional information loss caused by additional irreversible circuitry
compared to the irreversible time domain implementation. Our implementation
has no bit erasures whatsoever. The price that is paid reflects in
the larger number of gates used in the circuit: the number of gates
has only spatial consequences, information-related energy dissipation
is zero thanks to information conservation. Having no bit erasures
means theoretical decrease in energy consumption proportional to the
number of bits erased by the state of the art irreversible implementations\textendash multiplied
by the number of processing chains serving multiple transceivers,
it is clear that this quantity, as small as it may seem in the case
of a single transceiver, is indeed significant, especially in the
near future where the Landauer limit becomes the dominant bound in
semiconductor component power dissipation. As the circuits in our
solution perform arithmetical and logical operations in the same vein
as irreversible circuits in the state of the art implementations,
the performance of the two solutions is the same in terms of results
(the components are validated at the level of logical circuits).

On the other hand, in Fig. \ref{graphs}(b) we see the information
loss in the analog part of the system, and we differentiate two typical
environments, the chaotic cavity and the complex medium. The chaotic
cavity is an ergodic space with sensitive dependence on initial conditions
for waves. In such an environment there is little to no loss in the
information if the waiting time is long enough and the ADC resolution
is high enough. The complex medium is one with a large number of scatterers;
In such media, the difference is caused by some of the wave components
being reflected backwards by the scattering environment, hence not
reaching the TRM. Again, more information is retained with the increase
in the ADC resolution. However, as reported in \cite{derode_ultrasonic_1999},
the information loss from low-resolution ADC use does not affect the
performance of the algorithm. The analog part of the scheme remains
a topic of our future work, as it leaves space for improvements of
the scheme.

When the first prototype of a reversible FFT chip was introduced \cite{skoneczny_reversible_2008},
it was shown that the implementation based on 8-bit adders and 11-bit
multipliers requires 40,000 transistors. As the idea of information
and energy conservation in computation allows for very dense packing
(no heat dissipation to limit the density), this order of magnitude
for our solution is acceptible (the rest of the circuitry we introduce
in the chain needs two or three orders of magnitude fewer transistors),
and if we opt for no frequency domain processing, i.e. just using
memory, we can perform the task with less than 1,000 transistors.

In this first application of reversible hardware to a physical process
we have demonstrated the complementary nature of wave reversal and
computational reversal which can be put to use. Our implementation
of wave time reversal using reversible hardware carries the promise
of reduced power consumption, and it also fits in the bigger picture:
in this vision of the future, all computation is reversible, no matter
if it is performed on classical or quantum basis.

This solution is just the first step in the proliferation of reversible
computation in communications. The inherent reversible properties
of communications, including but not limited to channel reciprocity
and transmitter-receiver duality, make communications technology an
area with a lot of potential in reversible computation. With the inevitable
penetration of quantum-based techniques in communications, this link
with reversible computation grows stronger and needs to be thoroughly
investigated.

\bibliographystyle{IEEEtran}

\vspace{-0.5in}

\begin{IEEEbiographynophoto}{Harun Siljak}
 (M '15) graduated from Automatic Control and Electronics Department,
University of Sarajevo (BoE 2010, MoE 2012) and International Burch
University Sarajevo (PhD 2015). Currently he is a postdoctoral Marie
Curie Fellow at CONNECT Centre, Trinity College Dublin, working on
reversible computation, complex and nonlinear dynamics in wireless
communications.
\end{IEEEbiographynophoto}

\vspace{-0.5in}

\begin{IEEEbiographynophoto}{Julien de Rosny}
 received the M.S. Degree and the Ph.D. degree from the University
UPMC, Paris, France in 1996 and 2000, respectively, in wave physics.
He was a postdoctoral researcher at Scripps Research Institute, California,
USA, in 2000-2001. In 2001, he joined CNRS at Laboratoire Ondes et
Acoustique, France. Since 2014, he is a CNRS senior scientist at Institut
Langevin, Paris, France. His research interests include telecommunications
in complex media, acoustic and electromagnetic waves based imaging.
\end{IEEEbiographynophoto}

\vspace{-0.5in}

\begin{IEEEbiographynophoto}{Mathias Fink}
 is the George Charpak Professor at ESPCI Paris where he founded
in 1990 the Laboratory \textquotedblleft Ondes et Acoustique\textquotedblright{}
that became in 2009 the Langevin Institute. He is member of the French
Academy of Science and of the National Academy of Technologies of
France. In 2008, he was elected at the College de France on the Chair
of Technological Innovation. His area of research is concerned with
the propagation of waves in complex media and the development of numerous
instruments based on this basic research. His current research interests
include time-reversal in physics, wave control in complex media, super-resolution,
metamaterials, multiwave imaging, geophysics and telecommunications.
He holds more than 70 patents, and has published more than 400 peer
reviewed papers and book chapters. 6 start-up companies with more
than 400 employees have been created from his research (Echosens,
Sensitive Object, Supersonic Imagine, Time Reversal Communications,
CardiaWave and GreenerWave).
\end{IEEEbiographynophoto}

\end{document}